\documentclass[conference]{IEEEtran}
\usepackage{blindtext, graphicx}
\usepackage{subcaption}
\usepackage{amsmath}
\let\labelindent\relax
\usepackage{enumitem}

\usepackage{algorithm}
\makeatletter
\renewcommand{\ALG@name}{Procedure}
\makeatother
\usepackage{algpseudocode}
\usepackage{lipsum}

\begin{document}

\parskip 2pt

\let\OLDthebibliography\thebibliography
\renewcommand\thebibliography[1]{
  \OLDthebibliography{#1}
  \setlength{\parskip}{0pt}
  \setlength{\itemsep}{0pt plus 0ex}
}

\algnewcommand\algorithmicswitch{\textbf{switch}}
\algnewcommand\algorithmiccase{\textbf{case}}
\algnewcommand\algorithmicassert{\texttt{assert}}
\algnewcommand\Assert[1]{\State \algorithmicassert(#1)}%
\newcommand{\thicktilde}[1]{\mathbf{\tilde{\text{$#1$}}}}
\algdef{SE}[SWITCH]{Switch}{EndSwitch}[1]{\algorithmicswitch\ #1\ \algorithmicdo}{\algorithmicend\ \algorithmicswitch}%
\algdef{SE}[CASE]{Case}{EndCase}[1]{\algorithmiccase\ #1}{\algorithmicend\ \algorithmiccase}%
\algtext*{EndSwitch}%
\algtext*{EndCase}%

\setlength{\abovedisplayskip}{3pt}
\setlength{\belowdisplayskip}{3pt}

\algnewcommand\algorithmicforeach{\textbf{for each}}
\algdef{S}[FOR]{ForEach}[1]{\algorithmicforeach\ #1\ \algorithmicdo}
\renewcommand{\algorithmicforall}{\textbf{every}}
\algtext*{EndWhile}
\algtext*{EndIf}
\algtext*{EndFor}

\title{Context-Aware Configuration and Management of WiFi Direct Groups for Real Opportunistic Networks\vspace{-1cm}}
\author{\IEEEauthorblockN{Valerio Arnaboldi, Mattia G. Campana, and Franca Delmastro}
\IEEEauthorblockA{IIT-CNR, Via G.Moruzzi, 1 56121, Pisa, ITALY\\
\{v.arnaboldi, m.campana, f.delmastro\}@iit.cnr.it}}

\maketitle

\begin{abstract}
Wi-Fi Direct is a promising technology for the support of device-to-device
communications (D2D) on commercial mobile devices. However, the standard as-it-is
is not sufficient to support the real deployment of networking solutions entirely based on
D2D such as opportunistic networks. In fact, WiFi Direct presents some characteristics that could limit the autonomous creation of D2D connections among users' personal devices. 
Specifically, the standard explicitly requires the user's authorization to establish a connection between two or more devices, and it provides a limited
support for inter-group communication.
In some cases, this might lead to the creation of isolated groups of nodes which cannot communicate among each other.
In this paper, we propose a novel middleware-layer protocol for the efficient configuration and management of WiFi Direct groups (WiFi Direct Group Manager, WFD-GM) to enable autonomous connections and inter-group communication. This enables opportunistic networks in real conditions (e.g., variable mobility and network size).
WFD-GM defines a context function that takes into account heterogeneous parameters for the creation of the best group configuration in a specific time window, including an index of nodes' stability and power levels.
We evaluate the protocol performances by simulating three reference scenarios including different mobility models, geographical areas and number of nodes. 
Simulations are also supported by experimental results related to the evaluation in a real testbed of the involved context parameters. We compare WFD-GM with the state-of-the-art solutions and we show that it performs significantly better than a Baseline approach in scenarios with medium/low mobility, and it is comparable with it in case of high mobility, without introducing additional overhead.
\end{abstract}


\IEEEpeerreviewmaketitle

\vspace{-0.2cm}
\section{Introduction}

The number of active mobile devices recently bypassed the world population, according to GSMA Intelligence\footnote{https://www.gsmaintelligence.com/}.
People often carry multiple devices in their pockets, each equipped with several wireless interfaces.
This generates a number of opportunities for the users to create wireless communications. Currently, wireless interfaces are mainly used to access the Internet through fixed infrastructures (WiFi access points or cellular base stations).
Despite this, many interfaces support also direct communication between devices (device-to-device communication, or D2D).
The support for D2D on commercial mobile devices such as smartphones, tablets and laptops attracted the interest of researchers in the field of mobile and pervasive computing~\cite{conti2012looking}.
In fact, it can foster the creation of distributed services that run on mobile devices and rely on the network formed by multiple direct connections between them to coordinate operations and to disseminate user-generated contents, without the need for a fixed infrastructure.

In this context, since devices follow people movements, the structure of the network is usually unstable, and nodes (or groups of nodes) might end up being temporarily isolated from the rest of the network.
To enable networking functions in these conditions (e.g., routing and content dissemination), several protocols have been proposed in the literature~\cite{pelusi2006opportunistic} based on the \textit{store-carry-forward} paradigm and dealing with variable delays during message propagation. 
These protocols paved the way for the definition of new context- and social-aware distributed
services and applications based on D2D, such as Mobile Social Networks (MSN)~\cite{delmastro2016people}, and novel mobile-based recommendation systems for content dissemination~\cite{arnaboldi2016personalized}~\cite{lo2010folksonomy}. 

However, to effectively deploy opportunistic networks in a real environment, through available commercial mobile devices, we must face with technical constraints 
introduced by the available communication standards like Bluetooth, WiFi, WiFi Direct (hereinafter WFD), NFC, and their implementation on mobile operating systems.
NFC interfaces have a very limited range ($<$ 20cm), which results in the need for
the users to put the devices in physical contact to create a connection between
them, and this is not reasonable for opportunistic networks. On the other hand, 802.11 standard originally provided an ``ad hoc'' mode, in which devices could communicate directly with each other in a peer-to-peer manner, but this mode has been explicitly removed by Android (unless the device is rooted) and iOS.
Commercial devices typically support the \emph{tethering} mode, through which a device can act as a hotspot Access Point (AP) in order to share its Internet access.
In this case, each node has a fixed role as AP or client, and two nodes with the same role cannot communicate. In addition, it has been demonstrated that the AP mode heavily consumes device's
energy~\cite{sharma2009cool}~\cite{keshav2012energy}~\cite{jung2014energy} since IEEE 802.11 standard does not include any power saving mechanism for the AP (assuming it as a continuously powered device).
D2D communications are also available through Bluetooth and WFD, which both introduce power saving techniques.
However, they require the explicit authorization by the user for each connection establishment. Specifically, Bluetooth pairing requires the selection of a pin, while WFD asks for the acceptance of a pop-up notification during the connection phase. 
These features limit the creation of autonomous connections and the deployment of opportunistic networks.
Moreover, both standards do not support communication among groups, even if these are in proximity.
This prevents the content dissemination in the network, maintaining the groups isolated.
 
In this paper, we propose a novel middleware-layer protocol for the efficient configuration and management of WFD groups (WFD Group Manager, WFD-GM) to enable the creation of opportunistic networks in real conditions. 
WFD-GM exploits the main features of WFD standard to discover devices in proximity, and then it exchanges context information among nodes in order to compute (in a distributed manner) a context function defined to identify the best group configuration in a specific time window.
It is designed for Android commercial devices, since it represents one of the most diffused mobile operating system and the most open to third-party development.

Specifically, WFD-GM combines two mechanisms of WFD standard. As a first step, it uses the {\it Service Discovery} function designed to support zero-configuration networking protocols (e.g., 
UPnP\footnote{https://openconnectivity.org/resources/specifications/upnp/specifications}
and Bonjour\footnote{https://support.apple.com/bonjour}) on top of Wi-Fi
connections. In this procedure, nodes are able to exchange the SSID and key of the group they belong to. 
This allows devices to avoid the manual user authorization for D2D connections and they can autonomously connect to each other. Even though this procedure overcomes the security level introduced by the mobile operating system, it operates at a first authentication level. We will show in the next section how it is possible to maintain additional security levels in opportunistic networks even implementing a simple key exchange procedure during the service discovery, as presented in \cite{wu2015security}.


WFD-GM includes also the definition of a context function that takes into account heterogeneous features of the devices (e.g., battery level, list of neighbors) in order to identify the best group configuration.
In fact, group configuration and establishment in WFD requires the identification of a node that assumes the role of `Group Owner' (GO), mainly acting as AP for the group, and the others acting as clients.
A node can become GO in an autonomous way (i.e., directly creating its own group) or by a negotiation phase between two devices in proximity.

However, selecting and creating the best group configuration is not always sufficient to guarantee an optimized network coverage and content dissemination.
In fact, by using only this initial procedure, nodes reside in the same group until they are in proximity and/or they have not consumed their resources.
In this case, the network might be formed of several isolated groups.
For this reason, we introduced an additional procedure in WFD-GM for selected nodes that are in the communication range of two or more separated groups.
Specifically, it can force the disconnection of a client from the original group and its subsequent connection to another group in proximity, making that node a traveler between the two groups and contributing to disseminate contents in the network.
On the other hand, a GO can decide to merge its group with others in proximity through a specific procedure.

The rest of the paper is structured as follows. Section~\ref{sec:wifidirect} describes the main Wi-Fi Direct operations. Section~\ref{sec:related_work} provides an overview of the existing research work in this field. Section~\ref{sec:our_solution} presents the details of WFD-GM, and in Section~\ref{sec:experiments} we present the evaluation metrics and the experimental results obtained in three realistic scenarios. Finally, in Section~\ref{sec:conclusions} we draw our conclusions and present directions for future works.

\vspace{-0.1cm}
\section{Wi-Fi Direct}
\label{sec:wifidirect}

WFD is based on the definition of \emph{P2P groups}, in which one device (called \emph{Group Owner} or simply \emph{GO}) implements the functionalities of a IEEE 802.11 AP and the others act as clients. In addition, WFD implements power saving services running on the GO in favor of its clients and the GO is in charge of running a DHCP to assign IP addresses to the clients to enable the communication~\cite{peer2016technical}. 
The clients of a P2P group can be both P2P-enabled or legacy devices
(i.e., not supporting WFD). In the latter case, clients cannot exploit
the enhanced features of WFD, but they may join the P2P group by connecting to the GO as they typically do with a traditional AP.
WFD allows devices to establish a P2P group through three different procedures: i) \emph{Standard}, in case two or more nodes discover each other and there is a negotiation phase for the GO election; ii) \emph{Autonomous}, when a device autonomously decides to create a group and becomes the GO, announcing itself through beacon messages; and iii) \emph{Persistent}, in case the devices use stored configuration parameters of a previous group to re-establish the same group and speed up the process.

Each of the three procedures exploits the main functionalities of WFD. Specifically, they mainly rely on (i)  \emph{Peer} (optionally \emph{Service}) \emph{Discovery}, (ii) \emph{GO Negotiation}, and (iii) \emph{WPS Provisioning}. In the following, we briefly describe the main characteristics of these features, highlighting their advantages and drawbacks in supporting a real opportunistic network. 
\vspace{-0.1cm}
\begin{description}[style=unboxed,leftmargin=0cm]
	\item[Peer Discovery]\hfill\\ In order to create a communication group,
          two P2P devices must first discover each other.  The Peer Discovery
          phase usually starts with a traditional 802.11 Wi-Fi scan,
          through which the devices are able to find existent P2P groups and
          traditional WLAN networks.  After this scan, the following discovery
          algorithm is executed.  First, a P2P device randomly selects one of
          the so called \emph{Social channels} (i.e., channel 1, 6, and 11 in
          the 2.4 GHz band) as its own \emph{Listen Channel}, i.e., the channel
          on which it will ``listen'' for discovery messages coming from other
          devices.  The chosen Listen Channel remains the same until the Peer
          Discovery is completed.  Then, the device continuously switches between two
          operative states: \emph{search} and \emph{listen}.  When it is in the
          former state, the device sends \emph{Probe Request} messages to each
          of the Social channels; instead, when it is in the latter one, the
          device listens for Probe Requests in its Listen Channel in order to
          respond with \emph{Probe Response} messages.  Finally, two devices
          discover each other when they are on the same channel, but in
          different discovery state.  Convergence of two devices on the same
          channel is assisted by randomizing the time spend in each
          state. Typically, this time is randomly distributed between 100 ms and
          300 ms~\cite{camps2013device}, but the actual amount of time is
          implementation dependent.\\

	\vspace{-0.1cm}
	\item[GO Negotiation]\hfill\\ Once two devices have discovered each other, they proceed with the Standard group formation, where the GO Negotiation procedure begins. This phase implements a three-way handshake used to agree on which device shall become GO, and the channel to be used for the communication.

	During the negotiation, nodes exchange a \emph{GO Intent} (GI): an integer value (from 1 to 15) with which a device expresses its willingness to act as GO. The device which sends the higher intent becomes the owner of the group. In order to prevent conflicts during the GO election (e.g., if two devices send the same GO Intent), a \emph{Tie breaker} bit is randomly
	set to 0 or 1 every time a GO Negotiation Request is sent. The device with the Tie breaker bit set to 1 will be elected as GO.

	Generally, the GI value is not related to the actual suitability of a node to act as GO. In Android, upper-layer applications can specify a GI, otherwise the WFD framework simply sets it with a random value\footnote{https://developer.android.com/guide/topics/connectivity/wifip2p.html}.\\
    

	\vspace{-0.5cm}
	\item[Service Discovery]\hfill\\ The Service Discovery is a WFD optional feature and it represents an extension of the Peer Discovery. In fact, it adds a message exchange phase among nodes in proximity by exploiting the Generic Advertisement Service (GAS) protocol defined in IEEE 802.11~\cite{ieee2012802}. GAS is a link layer query/response protocol that allows two non-associated 802.11 devices to exchange queries coming from a higher layer protocol (e.g., Bonjour or UPnP).  When a requester discovers another peer, it transmits one or more \emph{GAS Initial Request} frames, and the target responds with one or more \emph{GAS Initial Response} frames if it exposes some services. This procedure can be performed both as a complete discovery procedure, collecting additional information for the GO selection, and after the group formation to periodically check devices in proximity and to dynamically manage groups.
	
	According to \cite{experimenting-Delmastro}, Peer Discovery and GO Negotiation phases require several seconds, especially in the Standard procedure, introducing thus a not negligible delay in the group formation.
    For this reason, in WFD-GM we decided to avoid the GO Negotiation phase, and we exploit the Service Discovery procedure to exchange context
	information among devices related to single nodes' characteristics (e.g., available computational resources, or the battery status). This information is then used by each node to evaluate its suitability to become GO of the group as the result of a context function. In fact, one of the main targets of our protocol is to select the best GO in the surroundings in order to establish a stable and long lasting communication group, in addition to speed up the group formation process.
	Then, once the group is created, WFD-GM performs a periodic Service Discovery procedure and dynamically evaluates the group configuration, depending on the information shared by surrounding devices, activating, in case they are needed, traveling or merge operations.
    
	In addition, in order to make the group formation as much autonomous as possible, WFD-GM also avoids the explicit user's authorization for D2D communication. To this aim, it acts on the WPS Provisioning phase, as described below. \\
    
    \item[WPS Provisioning]\hfill\\ The main purpose of this phase is to establish a secure connection between the GO and the group members, after the explicit user's authorization (through a PIN confirmation or an Accept button).
	WFD implements thus the Wi-Fi Simple Configuration (WPS)~\cite{alliance2011wi} protocol, by requiring that the GO generates and issue the network credentials to its clients. 
	WPS uses WPA-2 with a randomly generated Pre-Shared Key (PSK) as security measure to protect the connections, and the Advanced Encryption Standard (AES)-CCMP in order to encrypt the transmissions.
	In this case, the user's authorization mainly focuses on the authentication process of the connecting device, often ignoring the reason of the connection request. If we consider a middleware framework designed to support the establishment of opportunistic networks, and the users willing to participate to the network through their mobile devices, we can also envision that the framework could obtain a general user authorization during the installation phase and autonomously manage the devices' connection while maintaining the data encryption mechanism. Then, additional security measures could be defined at the upper layers to implement secure routing mechanisms, trust management and cooperation protocols, and application/user specific privacy protections. This approach has been largely studied in the literature and presented in \cite{wu2015security}, highlighting the different levels of security we can implement in an opportunistic network while supporting the autonomous generation and management of groups of nodes.
	
	Following this approach, WFD-GM exploits the Service Discovery procedure, running after the group creation,  to exchange the encrypted network credentials of the GO with the nodes' in proximity. In this way, those nodes can autonomously join the WFD group as legacy clients, without any user intervention to authorize the connection.
	Then, upper-layer security mechanisms (which are currently not provided by WFD-GM) can be implemented to guarantee an efficient access control and trust operations among nodes running the same middleware framework and WFD-GM protocol. 

\end{description}

\vspace{-0.4cm}
\section{Related Work}
\label{sec:related_work}

In the last few years, WiFi Direct has generated a lot of interest in the opportunistic networking research community.
However, most of the works in the literature focused on the experimental evaluation of basic standard features (with a limited number of nodes), trying to overcome WFD limitations through hacks and/or by rooting the devices. 
We can also divide related work depending on their main optimization target: (i) selection of the best GO, (ii) autonomous group formation (bypassing the user's authorization), and (iii) inter-group communication.

\vspace{-0.3cm}
\subsection{Selection of the best GO}

As previously described, GO selection represents one of the most important phases of the entire protocol since a ``good'' GO can be able to guarantee  a communication path among the highest number of nodes in proximity, and to improve the entire group performance. 
WD2~\cite{zhang2014wd2} is an algorithm aimed at automatically
selecting a GO based on the Received Signal Strength Indication (RSSI)
measurements. In this case, each device collects the RSSI reading from nearby devices, and a GO Intent (GI) value is calculated based on such collected measurements.
The devices then exchange their GI values during a modified
discovery phase. The device that exposes the highest GI value creates the group.
WD2 has been validated on simple network topologies composed by a maximum of five Android devices, and it effectively speeds up the standard Android group formation. However, it requires a modified implementation of WFD native framework, which limits the applicability of the algorithm in real scenarios.

Other researchers propose more advanced strategies for the selection of the best GO candidate. For Menegato et al.~\cite{botrel2014dynamic}, the
device who act as GO should change dynamically, and the choice of a new GO
should be based on the residual energy of the candidates.
In~\cite{laha2015energy}, the authors proposed three different approaches to
choose the GO: i) the device with the higest ID in the surroundings, ii) the peer that
has the shortest average distance from the other nodes, iii) the node with less mobility with respect to its neighbors.
However, considering only a feature at a time could be not sufficient to manage the complex dynamic that typically govern a real
mobile scenario. For instance, the device that discovers the highest number of
neighbors might also be the one with the lowest battery level. Selecting it as the GO
would lead to a more extended group, but probably characterized by a very limited duration.
WFD-GM leverages a combination of several features to evaluate the suitability of a node to act as GO in a specific context, as described in Section~\ref{sec:our_solution}.

\vspace{-0.2cm}
\subsection{Autonomous group formation}

Once a GO is selected, the other peers must connect to it in order to start the
communication. As described in Section~\ref{sec:wifidirect}, the WPS
Provisioning phase might represent a limitation to the use of WFD in
mobile scenarios. In the literature, researchers proposed different approaches
in order to allow users' personal devices to autonomously form WFD
groups (i.e., without asking for the explicit user's authorization). 
Wong et al~\cite{wong2014automatic} have been probably the first to tackle this problem. They
exploited WFD ability to support legacy devices and the Service
Discovery in order to avoid user intervention in the formation of groups.  The
device that elects itself as GO, sends the security credentials of the group to
the other peers through a Service Discovery Response message. In this way, the
peers in proximity can connect to it as legacy clients without the need of the user's intervention. This solution exploits the same approach that we adopted in WFD-GM to avoid the explicit user's authorization but it does not take into account the other two fundamental features: (i) the best GO selection and (ii) the communication among isolated groups. In addition, it does not take into account security issues derived from previous operations.
The solution proposed in~\cite{shahin2015efficient} exploits the same approach and includes a simple criterium to elect the GO: the best candidate is the node with the highest battery level among those in proximity. In addition, exploiting the ability of WFD to support legacy devices, this solution also enables inter-group communication, but it requests a customization of the native WFD framework.


\vspace{-0.2cm}
\subsection{Inter-group communications}

On current WFD implementations, especially on Android OS, each group is characterized by the same IP subnet. Thus, even if different groups could be interconnected, due to the presence of some nodes in both communication ranges, this is not possible~\cite{shahin2016ip}. 
Some recent work attempt to bypass the IP subnet constraint in different ways. Specifically, Casetti et al.~\cite{casetti2015content} proposed a solution that allows a GO to manage a group and, at the same time, to connect itself as a legacy client to a second group. The system exploits a combination of unicast and broadcast messages in order to transmit data among different groups, introducing however a non-negligible overhead in the overall communication.
A recent work~\cite{shahin2016ip} proposes an algorithm which exploits the
Service Discovery mechanism in order to allow devices to negotiate distinct IP
subnets before the establishment of the groups. Once the GOs agree on the IP
subnets, each of them creates its own group and uses its proposed IP
subnet. However, the solution is based on a customization of the Android WFD framework implementation to force the replacement of the default fixed IP
subnet with the negotiated one. This limits the applicability of the solution on a broader set of devices, and consequently does not allow large-scale deployment of opportunistic networks.

\vspace{-0.2cm}
\subsection{Other approaches}

Other solutions, such as~\cite{shahin2015alert} and~\cite{tsiridis2016adding}, embed application messages directly into Service Discovery frames. This kind of approach does not require any
infrastructure, connections, or groups formation for data exchange, relying only on the
service discovery announcements and requests to propagate messages between peers. Even though this approach could represent a valid solution to exchange small amount of data between devices (e.g., alerts or advertisements), it has a very limited bandwidth, which might not be sufficient for many real world situations.

\setlength{\textfloatsep}{.2cm}
\begin{algorithm}[t]
	\caption{Main procedure}
	\begin{algorithmic}[1]
		\footnotesize
		\Procedure{MAIN}{\null}
			\State $\left< mSSID, mPasskey \right > \gets$ createGroup()
			\State $mGO \gets ME$, $\mathrm G_M \gets \emptyset$, $\mathrm L_B \gets \emptyset$, $L_N \gets \emptyset$, $state \gets GO1$
	  		\State ServiceDiscovery(); UpdateStabilityIndex(); UpdateMyServiceInfo();
			\State MessageReceiver(); EventReceiver()
      \ForAll{$T_d$ seconds}:                               \Comment{Main loop}

	  	\Switch{$state$}
		  \Case{\textbf{GO1}}: \texttt{GoElection()} \EndCase
		  \Case{\textbf{GO2}}: \texttt{DisbandGroup()} \EndCase
		  \Case{\textbf{GO3}}: \texttt{EvalMerge()} \EndCase
		  \Case{\textbf{C1}}: \texttt{EvalTraveling()} \EndCase 
		\EndSwitch





      \EndFor
		\EndProcedure
	\end{algorithmic}
  \label{algo:main}
\end{algorithm}

\vspace{-0.3cm}
\section{WiFi Direct Group Manager}
\label{sec:our_solution}

WFD-GM combines all the operations sketched in Section~\ref{sec:wifidirect}, implemented through Android SDK version 14-25. The protocol, as detailed in Procedures~\ref{algo:main},~\ref{algo:traveler}, and~\ref{algo:merge}, runs on each single device.
In order to minimize the time required for a group formation and to optimize the credentials exchange, WFD-GM starts on each node by creating a WFD group in which the local device autonomously becomes GO (initially without any associated client).
In Android, this operation is performed through the \texttt{createGroup()} API, which also automatically generates the SSID and the group credentials (i.e., the WPA2-PSK key). 
This information is then included in the Service Discovery frames to allow autonomous connections.

After this operation, five parallel procedures start and keep running until the termination of the protocol.
The first procedure is the \texttt{ServiceDiscovery}, which
performs a continuous WFD Service Discovery and maintains an updated list ($L_N$) of the devices in proximity. The messages exchanged during the Service Discovery include, in addition to the group's credentials, an index of the local node suitability to become/remain GO of a larger group.
We define $s(ln)$  (called \textit{Suitability index})  as a function of the following set of context features: i) $r_{ln}$, the amount of available resources of the local device (e.g., battery level, free CPU, free memory), ii) $pp_{ln}$, the current number of peers discovered in proximity, iii) $c_{ln}$, the capacity of the node (i.e., the number of incoming connections that the device can still accept), and iv) $st_{ln}$, the \emph{stability index}, which provide a measure of the ability of the node to create a long lasting WFD group (i.e., a group that will not be rapidly destroyed due to the local node's mobility). 
More formally:
\begin{equation}
s(ln) = \omega_1 \cdot r_{ln} + \omega_2 \cdot pp_{ln} + \omega_3 \cdot c_{ln} + \omega_4 \cdot st_{ln},
\label{eq:goodness}
\end{equation}
where the weights $\omega_{1,\cdots,4}$ govern the relative importance of each feature in the overall computation of $s(ln)$.

The stability index $st_{ln}$ evaluates both the mobility of the local node and how much its surrounding environment changes over time.
Currently, we consider it as a function of the nodes in proximity ($L_N$), but more complex approaches can be taken into account (e.g., a function of the geographical locations visited by the node in the past). The \texttt{UpdateStabilityIndex} procedure is in charge to update $st_{ln}$ every $T_{st}$ seconds as follows.
Every time $L_N$ changes, it calculates the difference between the current list of neighbors and the one of the previous time window, then computing the Jaccard index of the two lists.
Then, it updates a running average $\bar{J}$ of the Jaccard indices calculated since the last update of $st_{ln}$. Finally, the stability index is updated with the following formula:
\begin{equation}
st_{ln} = st'_{ln} \cdot \omega_{st}^1 + \bar{J} \cdot \omega_{st}^2,
\label{eq:stability}
\end{equation}
where $st'_{ln}$ is the stability index calculated in the previous time window of $T_{st}$ seconds, and the weights $\omega_{st}^1$ and $\omega_{st}^2$ govern the relative importance of the past stability index in the current computation.
The \texttt{UpdateMyServiceInfo} procedure is in charge of updating the information included in the Service Discovery, and to this aim it uses the last updated stability index $st_{ln}$.

In order to manage the group dynamically, reflecting nodes' mobility, the protocol defines two asynchronous procedures running concurrently to the previous ones.
The \texttt{EventReceiver} procedure constantly listens for incoming connection requests from other devices and for clients' disconnections from the local group, maintaining an updated list of the group members ($G_M$). This events are only managed by GO nodes and, after each event, they broadcast a \texttt{GROUP\_INFO} message to all their clients containing the updated $G_M$ list, allowing them to maintain an updated view of the group. Note that WFD does not provide this feature natively. The second procedure is the \texttt{MessageReceiver}, which is in charge of receiving and processing incoming control messages depending on the local node status.

\begin{figure}[t]
	\centering
    \includegraphics[width=.31\textwidth]{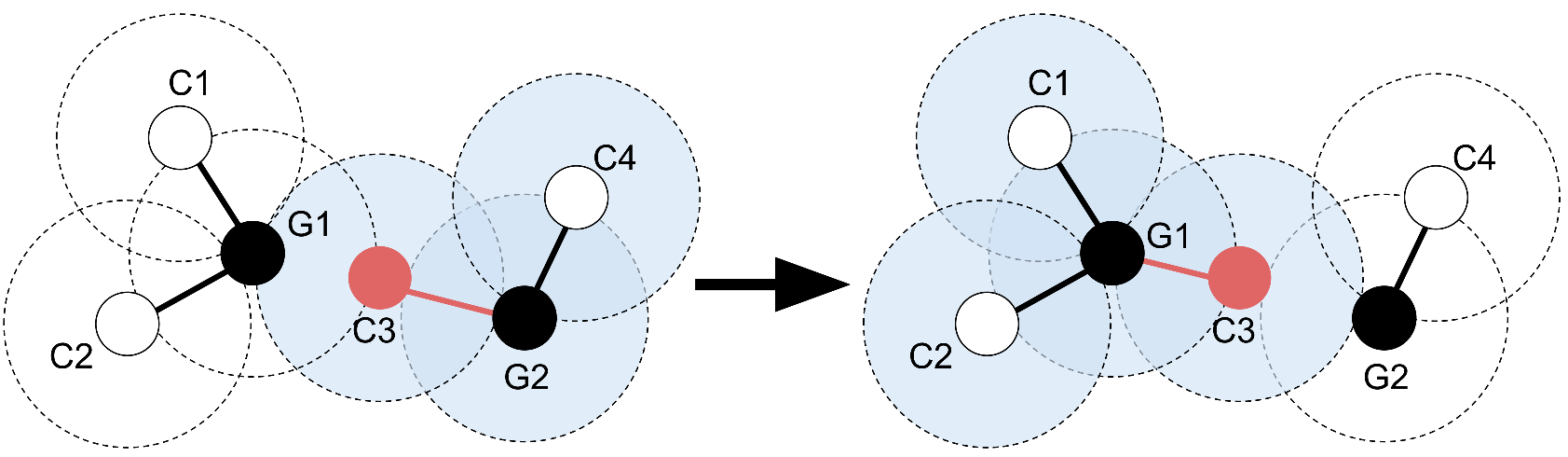}
    \caption{Traveling between two groups.}
    \label{fig:traveler}
	\vspace{-.88cm}
\end{figure}
\setlength{\textfloatsep}{1pt}
\begin{algorithm}[t]
	\footnotesize
	\caption{Evaluate Travelling}
	 \label{algo:traveler}
	\begin{algorithmic}[1]
		\footnotesize
		\Procedure{EvalTraveling}{}
			\State $r = rand(0, 1)$
			\If {$r \leq p_T$}
				\State $\mathrm L_B = \mathrm L_B \cup \{mGO, T_{Btravel}\}$ and Disconnect($mGO$)
			\EndIf
		\EndProcedure
	\end{algorithmic}
\end{algorithm}

\begin{figure}[t]
	\centering
    \includegraphics[width=.3\textwidth]{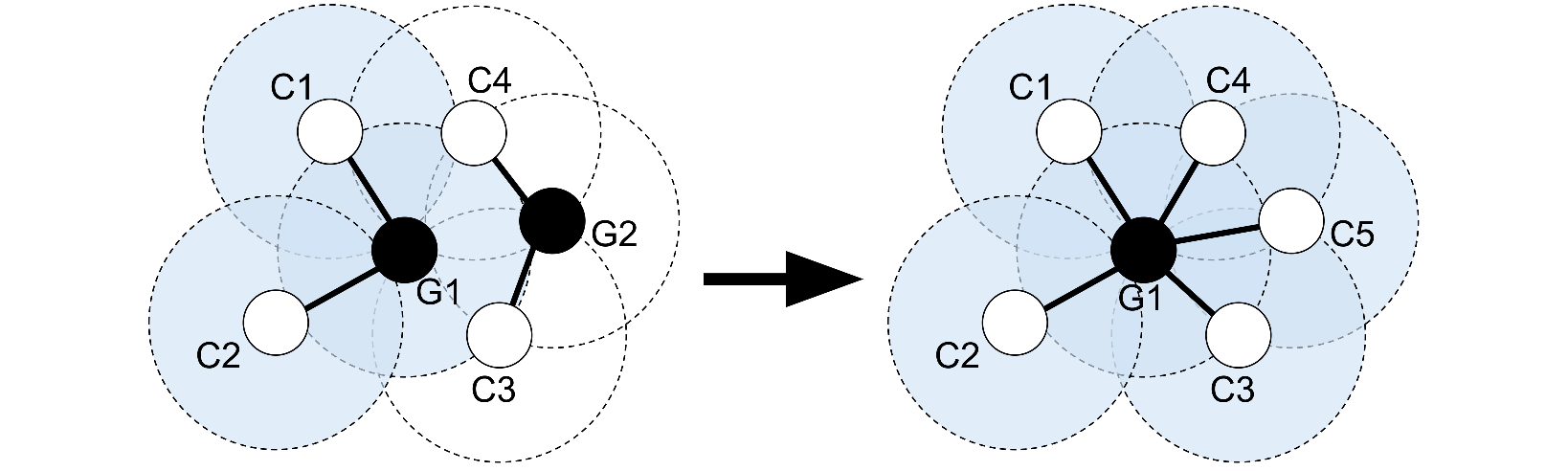}
    \caption{Merge of two isolated groups.}
    \label{fig:merge}
	\vspace{-.56cm}
\end{figure}
\setlength{\textfloatsep}{1pt}
\begin{algorithm}[t]
	\footnotesize
	\caption{Evaluate Merge}
	 \label{algo:merge}
	\begin{algorithmic}[1]
		\footnotesize
		\Procedure{EvalMerge}{}
			\State $g_{best}$ = best\_go($GO_N$)
			\If {$g_{best} != ME$}
				\State Send VISIBILITY\_REQ($g_{best}$) to the clients
				\State $V_R = $ \textbf{wait} VISIBILITY\_RESP from the clients
				\State $t = \left| \{r_i \in V_R : r_i == true\} \right|$

				\If {$t \geq \left | G \right | + 1$}
					\State Send MERGE\_WARNING($g_{best}$) to the clients
					\State \texttt{DisbandGroup()} and Connect($g_{best}$)
				\EndIf
			\EndIf
		\EndProcedure
	\end{algorithmic}
\end{algorithm}

\begin{figure*}[t]
	\begin{subfigure}{.25\textwidth}
		\captionsetup{skip=0.02cm}
		\centering
		\includegraphics[width=\linewidth]{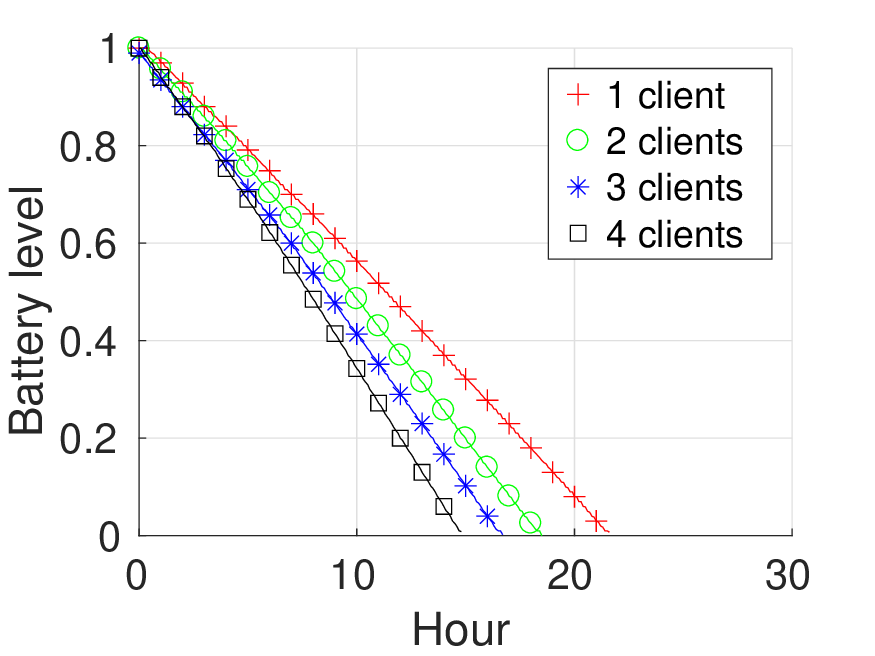}
		\caption{Measured GOs}
		\label{fig:start_go}
	\end{subfigure}%
	\begin{subfigure}{.25\textwidth}
		\captionsetup{skip=0.02cm}
		\centering
		\includegraphics[width=\linewidth]{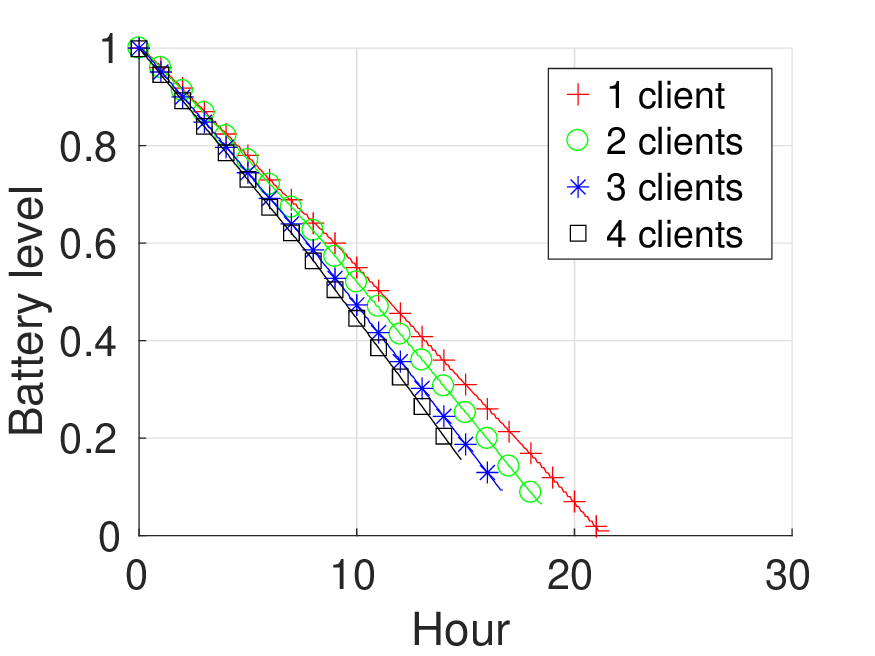}
		\caption{Measured Clients}
		\label{fig:start_clients}
	\end{subfigure}%
	\begin{subfigure}{.25\textwidth}
		\captionsetup{skip=0.02cm}
		\centering
		\includegraphics[width=\linewidth]{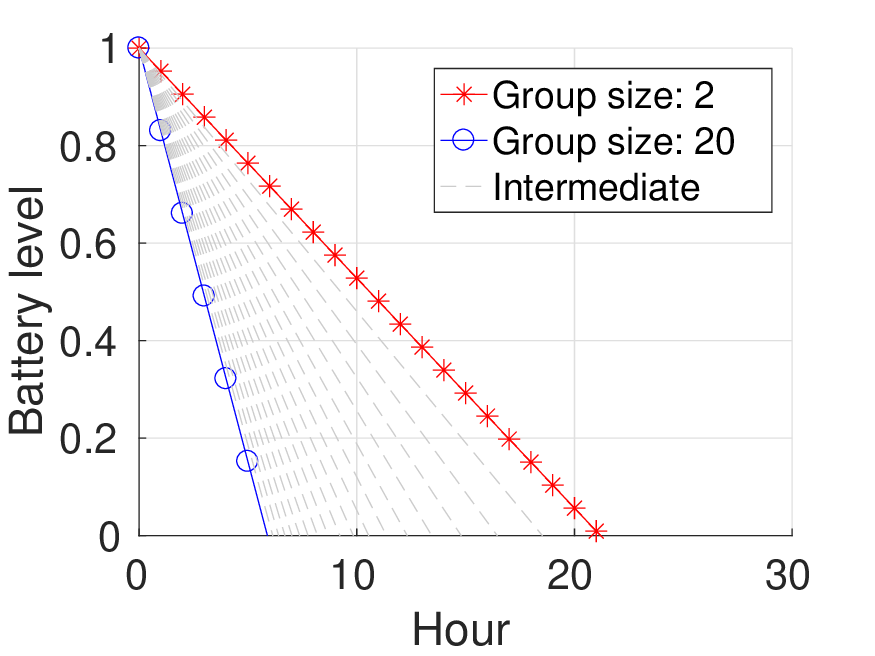}
		\caption{Predicted GOs}
		\label{fig:predicted_gos}
	\end{subfigure}%
	\begin{subfigure}{.25\textwidth}
		\captionsetup{skip=0.02cm}
		\centering
		\includegraphics[width=\linewidth]{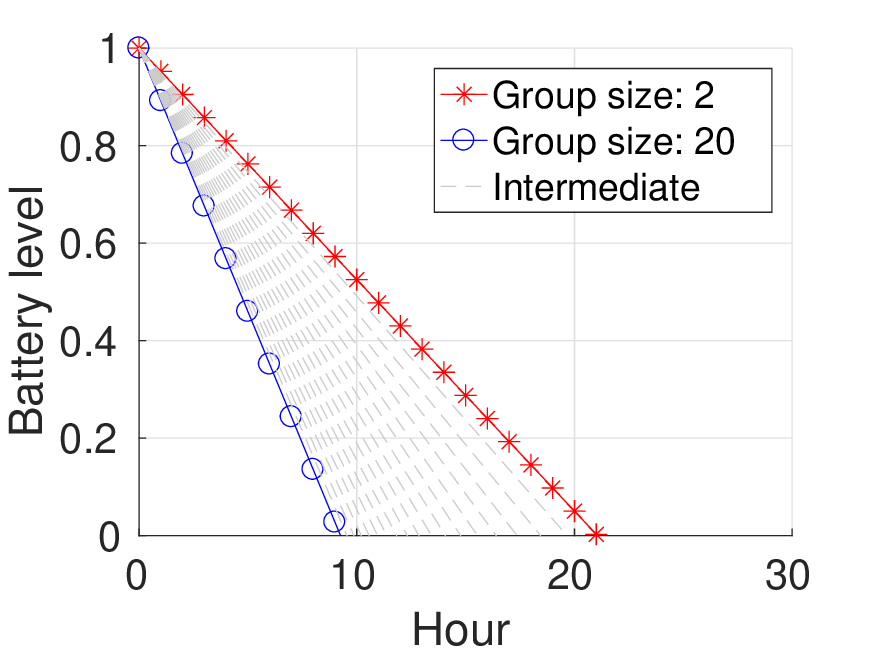}
		\caption{Predicted Clients}
		\label{fig:predicted_clients}
	\end{subfigure}
	\vspace{-.4cm}
	\caption{Battery depletion fitting.}
	\label{fig:fitting}
	\vspace{-0.8cm}
\end{figure*}

After launching these procedures, the protocol executes its main loop in which, every $T_D$ seconds, it checks the status of the local variables and the current role of the local node to choose and execute the appropriate action.
Specifically, if the local node is a GO, it can be in one of the following status:
\begin{description}[style=unboxed,leftmargin=0cm]
\item[GO1:]has no associated client ($G_M=\emptyset$), but the list of peers in proximity is not empty ($L_N \neq \emptyset$). The node must decide whether to remain GO or to connect as a client to another peer, using the \texttt{GoElection} procedure.
This compares the $s(ln)$ of the local node with those received by the others. If the local node has the highest  $s(ln)$, it remains GO and waits for incoming connections, otherwise it connects to the best GO as a client.
\item[GO2:]has some connected client ($G_M\neq \emptyset$) but the amount of resources consumed to manage the current group is beyond a predefined threshold $res_{th}$. The node sends a \texttt{GROUP\_BYE} message to its clients in order to alert them that it is destroying the group for limited resources. Then, it disbands the group and comes back to the inital status \textbf{GO1}.
\item [GO3:]has discovered other GOs in proximity ($GO_N$), with or without associated clients. It executes the \texttt{EvalMerge} procedure (Procedure~\ref{algo:merge} and Fig.~\ref{fig:merge}), aimed at evaluating the advantages of merging its local group with the others in proximity in order to form a larger group. The procedure firstly selects the best GO in proximity from the $GO_N$ list, based on the suitability index $s_{ln}$. If the best GO for the merge ($go_{best}$) is not the local node, this asks to its client if $go_{best}$ is in their respective proximity ranges and waits for the responses. Then, if the majority of the clients respond positively, the local node sends a \texttt{MERGE\_WARNING} message to its clients (to notify the merge decision), disbands the group and connects to $go_{best}$. Otherwise, it maintains its current status role of GO.
\end{description}


If the local node is a client, it can be only in the \textbf{C1} status and executes the \texttt{EvalTraveling} procedure (Procedure~\ref{algo:traveler} and Fig.~\ref{fig:traveler}).
Specifically, with probability $p_T$, inversely proportional to the group cardinality $\left|G_M\right|$, it disconnects from its current group, and places the GO in a blacklist ($L_B$) for a fixed amount of time ($T_{Btravel}$) to avoid considering it as potential GO during the subsequent \emph{GoElection} procedure. Finally, the local node returns to the \textbf{GO1} status in order to choose which group to connect among those in the $GO_N$ list.


A node assuming the role of client performs then additional actions depending on the type of message it receives from its current GO. Specifically, it can receive the following messages:
\begin{description}[style=unboxed,leftmargin=0cm]
\item[GROUP\_BYE:]means that the GO is disbanding the current group and a new one will be formed.
Therefore, it places the GO in $L_B$ for $T_B$ seconds, and it comes back to its initial GO status (without clients), \textbf{GO1}.
\item[VISIBILITY\_REQ($go_{best}$):]means that the current GO is evaluating a possibile merge operation and it selected the new best GO ($go_{best}$). The local node must verify its proximity to $go_{best}$ and appropriately reply to the current GO.
\item[MERGE\_WARNING:]means that the GO decided to disband its group in favour of a new GO, indicated as $go_{best}$ in the previous \texttt{VISIBILITY\_REQ} message. The local node places the GO in $L_B$ for $T_B$ seconds and it sends a connection request to the $go_{best}$ if in proximity, otherwise it comes back to its initial \textbf{GO1} status in order to execute the \texttt{GoElection} procedure.
\end{description}

WFD-GM is thus able to autonomously create connections between devices in proximity and to manage the creation of optimal groups (with respect to the $s(ln)$ value of the GOs). In addition, it allows inter-group communication through groups' merge operations and travelling clients.

\vspace{-0.1cm}
\section{Experimental evaluation}
\label{sec:experiments}

To evaluate WFD-GM performances and compare it with some reference solution, we decided to implement also a \emph{Baseline} protocol. It just implements the group's creation by using a simple rule to select the GO among nodes' in proximity: it chooses the one with the highest MAC address. Baseline executes the GO selection at the beginning of the protocol and the GO maintains its role until the end of its resources or in case it moves out of the connectivity range of all the group's members. It basically exploits WFD Service Discovery to exchange the group's credential to enable autonomous connections among nodes, and it does not implement any additional strategy for the group management (e.g., merge operations or traveling nodes). Therefore, Baseline is comparable with the state-of-the-art solutions presented in Section~\ref{sec:related_work}. 

We compare WFD-GM and Baseline in three simulation scenarios representing three real world use cases involving a variable number of nodes characterized by different mobility models.  In addition, since WFD-GM is characterized by a context function to evaluate the suitability of a node to assume or maintain the role of GO (Eq.~\ref{eq:goodness}), we need an estimation of the parameters to be included in the simulation set up. To this aim, we conducted a set of real experiments aimed at evaluating the resource consumption on real commercial devices related to the execution of main WFD operations (i.e., Service Discovery, Message exchange) and the capacity of a node in terms of maximum number of acceptable connections.
Therefore, in this section we present, firstly, the experimental results related to power consumption and groups' configuration. Then, we describe the characteristics of the simulation scenarios, the metrics used to compare the performance of Baseline and WFD-GM, and the achieved results.


\begin{figure*}[t]
	\centering
	\begin{subfigure}{.26\textwidth}
		\captionsetup{skip=0.1cm}
		\centering
		\includegraphics[width=\linewidth]{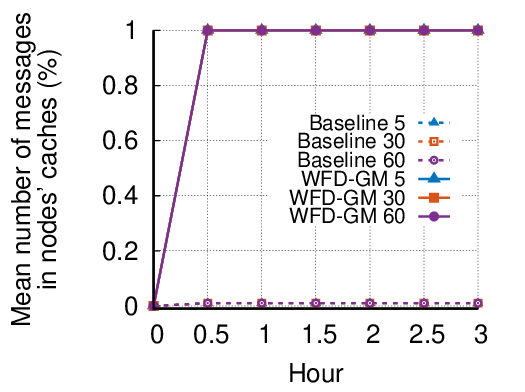}
		\caption{Concert}
		\label{fig:concert_messages}
	\end{subfigure}%
	\begin{subfigure}{.26\textwidth}
		\captionsetup{skip=0.1cm}
		\centering
		\includegraphics[width=\linewidth]{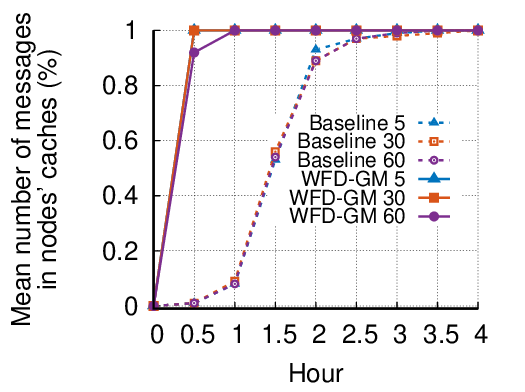}
		\caption{ComiCon}
		\label{fig:comicon_messages}
	\end{subfigure}
	\begin{subfigure}{.26\textwidth}
		\captionsetup{skip=0.1cm}
		\centering
		\includegraphics[width=\linewidth]{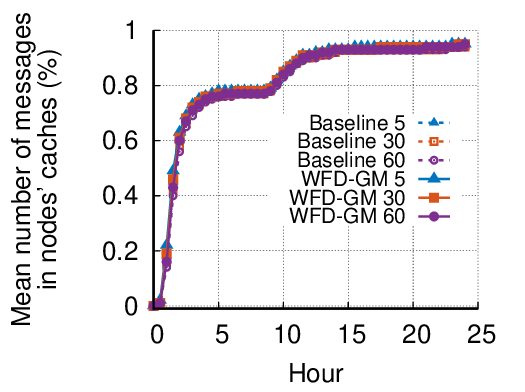}
		\caption{Helsinki}
		\label{fig:helsinki_messages}
	\end{subfigure}
		\vspace{-.4cm}
	\caption{Message diffusion performance}
	\label{fig:message_diffusion}
	\vspace{-.8cm}
\end{figure*}

\vspace{-0.2cm}
\subsection{Context parameters estimation}

The main purpose of WFD-GM is to allow mobile devices to autonomously create an opportunistic network and to dynamically manage its configuration according to nodes' mobility and with a fairly usage of the nodes' resources. Clearly, the most critical resource for a smartphone is represented by the battery consumption, which also represents a critical factor for the user experience while running a mobile app or framework on the personal mobile device.
We decided thus to model the power consumption on simulated nodes with respect to the main WFD operations implemented in our protocol in order to provide also an evaluation of the overhead introduced by WFD-GM on real devices.
To this aim, we carried out an experimental evaluation of the time required to entirely consume the battery of some commercial devices while assuming the role of GO or client in different configurations. We performed a series of empirical experiments with different smartphones (LG Nexus 5 and Motorola Nexus 6) equipped with different versions of Android (6.0.1 and 7.1.2).

In the first set of experiments, we considered a node assuming the role of GO without any associated client and we evaluated the power consumption while simply maintaining the GO status, and running a Service Discovery procedure every 2 minutes (the default time duration set in the Android P2P Framework\footnote{see the com.android.server.wifi.p2p.WifiP2pServiceImpl.java class in the Android P2P Framework source codes}). We repeated this experiment with an incremental number of peers in proximity (up to 10 devices).
We observed that there is no relevant difference between the two operations in terms of battery consumption. The overall measured cost is the same as maintaining the Wi-Fi interface active without performing any network connection or data transfer. In both experiments, the battery depletion is strictly linear, with a fall of approximately 20\% of the battery capacity every 5 hours. We used this data to update the battery level of each node during the simulation in case it performs the WFD Service Discovery procedure and it is not connected to any other peer.

Then, we performed a set of experiments to estimate the battery consumption of each node involved in a WFD group with a variable number of members (from 1 to 4 connected clients). The limit of 4 clients per group reflects the capacity of the commercial devices we used in our experiments (i.e., LG Nexus 5 and Motorola Nexus 6). In fact, we experienced that, for both models, when this limit is reached, the DHCP module running on the GO is not able to assign additional IP addresses to new clients and their connection request fails. The number of supported incoming connections strictly depends on the manufacturer's implementation, and it cannot be changed by the applications on not-rooted devices. In fact, we experienced different group cardinality with other commercial devices (e.g., up to 10 clients for a HTC Nexus 5X or Xiaomi Mi5 running as GO). Thus, to reproduce a realistic scenario, in which users are equipped with heterogeneous mobile devices, we assigned to each node in the simulation a \emph{capacity} parameter randomly chosen between 4 and 15 (i.e., the maximum number of acceptable incoming connections).

To estimate the battery consumption in a group configuration, we deployed a simple application in which devices create the group and each member constantly sends data to the others with a frequency of one message every 100 milliseconds. Even if the transmission frequency can be quite high for a mobile application use case, it lets us to model the battery depletion in a worst case scenario.
Figures~\ref{fig:start_go} and~\ref{fig:start_clients} respectively show the discharge curve of the battery on each single node.
We can note that the curves follow a linear trend, but the GO generally discharges faster then the clients.
This is due to the fact that, in a WFD group, the GO is also in charge of enabling communication between clients, forwarding all the messages exchanged by the members of its group.

To exploit these results in our simulation scenarios, we used a linear regression model to predict the battery consumption on a single node involved in larger groups. 
Figures~\ref{fig:predicted_gos} and~\ref{fig:predicted_clients} show the predicted discharging curves for both the GO and the clients in groups with up to 20 members.
Formally, the battery level at a certain time is given by the following linear function:
$b_l(t,n) = t \cdot (p_1 \cdot n - p_2) + 1$, where $t$ is the time in hour and $n$ is the number of clients in the group.
The values of $p_1$ and $p_2$ that we found by fitting real battery consumption data differ by the role played by the device in the group: $p_1 = -0.006802$ and $p_2 = - 0.03356$ if it is the GO, otherwise $p_1 = -0.003365$ and $p_2 = - 0.04075$ if it is a client.

\begin{figure*}[t]
	\centering
	\begin{subfigure}{.26\textwidth}
		\captionsetup{skip=0.1cm}
		\centering
		\includegraphics[width=\linewidth]{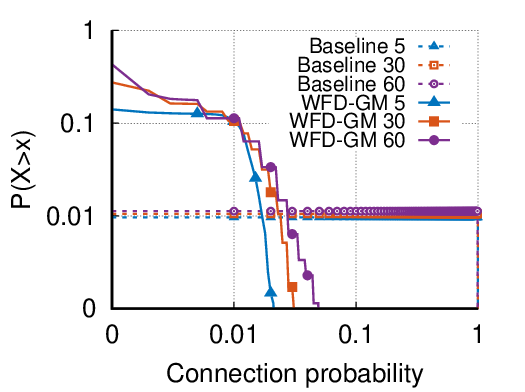}
		\caption{Concert}
		\label{fig:concert_ccdf}
	\end{subfigure}
	\begin{subfigure}{.26\textwidth}
		\captionsetup{skip=0.1cm}
		\centering
		\includegraphics[width=\linewidth]{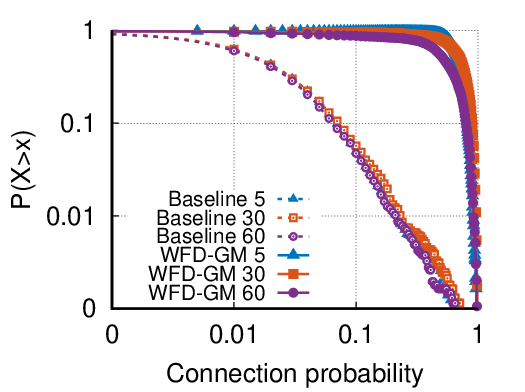}
		\caption{ComiCon}
		\label{fig:comicon_ccdf}
	\end{subfigure}
	\begin{subfigure}{.26\textwidth}
		\captionsetup{skip=0.1cm}
		\centering
		\includegraphics[width=\linewidth]{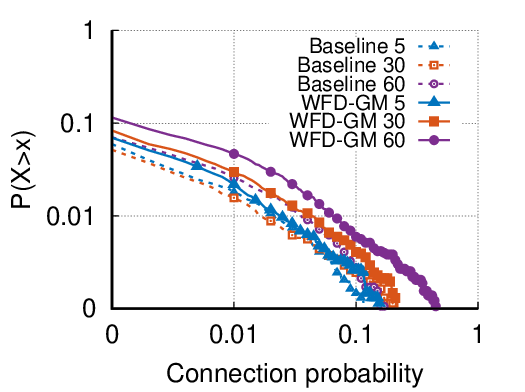}
		\caption{Helsinki}
		\label{fig:helsinki_ccdf}
	\end{subfigure}%
	\vspace{-.4cm}
	\caption{CCDF of the connection probability between nodes.}
	\label{fig:links_ccdf}
	\vspace{-.8cm}
\end{figure*}

\vspace{-0.2cm}
\subsection{Simulated use case scenarios}
We implemented three use case scenarios by using the ONE simulator~\cite{keranen-theone}.
Specifically, we envisioned three different application environments involving different numbers of users, with different mobility patterns and with a different geographical distribution: a concert, a convention venue and a working day in an European metropolitan city (Helsinki).

In the \emph{Concert} scenario, we replicated a medium-size concert with an audience of 1000 seated people, arranged in a 20 x 25 grid in an area of 500 m$^2$. We assumed that people are seated for the entire duration of the concert (i.e., 3 hours), without interruptions or users' movements during the exhibition (i.e., a static scenario). 
In the convention scenario (namely, \emph{ComiCon}), users are characterized by a moderate mobility. In this case the simulation lasts for 4 hours and 
the simulated geographical area is modelled as a grid of 4000mt x 2000mt. In such space, we distributed a total of 575 points of interest (POIs - e.g., exhibitors stands, toilets, eateries) in order to replicate the characteristics of a big convention of comics and games (e.g., the New York Comic Con\footnote{http://www.newyorkcomiccon.com}), and we simulated 2000 users moving following the \emph{ShortestPathMapBasedMovement} model implemented in ONE. This represents a map based movement model (i.e., the grid in this case) that uses Dijkstra's algorithm to find the shortest path between two random POIs. 
The simulation nodes are characterized by a speed in the range of $\left[ 0, 1.5 \right]$ m/s and each of them remains in a given POI for $t_w$ seconds, where $t_w$ is randomly drawn from $\left[ 600, 3600 \right]$ seconds. 
These parameters allowed us to model possible queues and crowds around the stands, characterizing the users with a moderate mobility, which is very common in a convention scenario.

Finally, we simulated an urban scenario (called \emph{Helsinki}), in which users are characterized by a high degree of mobility.
We used the Working Day Movement model~\cite{ekman2008working} implemented in ONE in order to simulate a typical working day of 4000 people.
The mobility model uses several highly customized mobility sub-models that define nodes' behavior during different daily activities in the Helsinki city center, such as staying at home, working, and evening activities with friends.
To simulate the movements between home and work, and between work and possible meeting points for evening activities, the model defines three additional mobility sub-models that are combined for each single node: car travel mobility, public transportation mobility, and walking mobility. For a complete description of the mobility model, the reader can refer to~\cite{ekman2008working}.
In this kind of scenarios, stable groups may be rare and the structure of the network configuration continuously evolves over time.

We expect that ComiCon and Concert scenarios highlight the advantages of using WFD-GM protocol since the limited nodes' mobility could create network's partitions, limiting the content dissemination among nodes of different groups. In addition, these crowded scenarios can benefit  from the autonomous generation and management of opportunistic networks, reducing the load of infrastructured wireless networks (characterized by limited capacity of APs or limited bandwidth).
On the other hand, in an urban scenario with high mobility, we expect that WFD-GM performs similarly to Baseline, while not introducing additional overhead.
In the following section we present the evaluation metrics and we discuss the experimental results.

\vspace{-0.2cm}
\subsection{Evaluation metrics and results discussion}

Baseline and WFD-GM share the same parameter $T_D$, which defines how frequently the two protocols take a decision, according to the status of the local node (Procedure~\ref{algo:main}). Therefore, we simulated each scenario for different values of $T_D$ (i.e., 5, 30, and 60 s), considering that the default value for a Service Discovery duration is 120s. In addition, we used the following values for the WFD-GM parameters: $\omega_{1,\ldots,4} = 0.25$ are the weights of the Suitability index (Eq.~\ref{eq:goodness}), $\omega_{st}^1 = 0.4$ and $\omega_{st}^2 = 0.6$ are used to compute the stability index (Eq.~\ref{eq:stability}), $res_{th} = 0.1$ is the resource threshold, and $T_B = T_{B\_travel} = 60$s are the blacklist times.

During each simulation, both Baseline and WFD-GM create a network of multi-hop paths among the nodes. This network can be represented as a graph, called \emph{Connectivity Graph}, in which two nodes are directly connected (1 hop) if they have participated in the same WFD group during the simulation. Formally, the \emph{Connectivity Graph} $CG = (V,E)$ is an undirected and weighted graph, where $V$ is the set of nodes, and each edge $e_{a,b} \in E$ between two nodes $a$ and $b$ is labeled with a weight representing the total connection time for two nodes (i.e., the sum of all the connection times between them).

In order to evaluate the $CG$ generated by the two protocols, we performed two different analysis. Firstly, we measured how fast a set of information can be disseminated within a given network configuration, assuming that nodes implement an epidemic forwarding algorithm. When a simulation starts, each node generates a message which contains the identification number of its creator. Every time a node joins a group, it sends all the messages contained in its own cache to each member of the group. Each node keeps in its own cache a copy of the received messages for the entire simulation. Every 30 minutes (simulation time) we measured the mean percentage of the messages contained in the nodes' caches.

Then, we performed a complex network analysis on $CG$ at the end of the simulation. Specifically, we evaluated the connection probability of each pair of nodes based on their total connection time, including also multi-hop paths, as an additional measure of the network connectivity. Then, we evaluated the partitioning degree of $CG$ in terms of number of connected components in the graph, and the percentage of nodes in the largest one (Tab.~\ref{tab:conn_components}). Both these measures highlight the degree of connectivity of the network depending on the nodes' mobility and the actions performed by the two protocols.

As a final measure, we compared the amount of nodes' resources used by the two protocols.In Tab.~\ref{tab:battery_stats}, we reported the mean value ($\mathbf{\mu}$), the median ($\tilde{\mathbf{x}}$), and the variance ($\mathbf{\sigma^2}$) of the nodes' battery level at the end of each simulation scenario, assuming $T_D = 30$s (we obtained similar results for the other values of $T_D$).Since Helsinki simulation is much longer than the other two scenarios (i.e., 24 hours), all the nodes discharged their batteries before the end of the simulations. Therefore, for this scenario we report the statistics about the (normalized) times used to expire the entire battery.

\begin{table}[b]
\centering
\renewcommand{\arraystretch}{1.3}
\caption{Number of $CG$'s connected components and percentage of nodes belonging to the largest one.}
\label{tab:conn_components}
\vspace{-0.2cm}
\begin{tabular}{c|r|r|r|r|r|r|}
\cline{2-7}
\multicolumn{1}{l|}{\textbf{}}      & \multicolumn{2}{c|}{\textbf{Helsinki}}                   & \multicolumn{2}{c|}{\textbf{ComiCon}}                    & \multicolumn{2}{c|}{\textbf{Concert}}                    \\ \hline
\multicolumn{1}{|c|}{\textbf{$\mathbf{T_D}$}} & \multicolumn{1}{c|}{Baseline} & \multicolumn{1}{c|}{WFD-GM} & \multicolumn{1}{c|}{Baseline} & \multicolumn{1}{c|}{WFD-GM} & \multicolumn{1}{c|}{Baseline} & \multicolumn{1}{c|}{WFD-GM} \\ \hline
\multicolumn{1}{|c|}{\textbf{5}}    & 8 (99\%)                      & 6 (99\%)                 & 1 (100\%)                     & 1 (100\%)                & 92 (2\%)                      & 1 (100\%)                \\ \hline
\multicolumn{1}{|c|}{\textbf{30}}   & 10 (99\%)                     & 9 (99\%)                 & 1 (100\%)                     & 1 (100\%)                & 95 (2\%)                      & 1 (100\%)                \\ \hline
\multicolumn{1}{|c|}{\textbf{60}}   & 11 (99\%)                     & 9 (99\%)                 & 1 (100\%)                     & 1 (100\%)                & 95 (2\%)                      & 1 (100\%)                \\ \hline
\end{tabular}
\end{table}

\begin{table}[t]
\renewcommand{\arraystretch}{1.3}
\centering
\vspace{-0.2cm}
\caption{Power consumption statistics}
\label{tab:battery_stats}
\vspace{-0.2cm}
\begin{tabular}{l|c|r|r|c|r|r|c|r|r|}
\cline{2-10}
& \multicolumn{6}{c||}{\textbf{Final battery level}}  	& \multicolumn{3}{c|}{\textbf{Time of discharge}}          	\\ \cline{2-10} 
\textbf{}  & \multicolumn{3}{c|}{\textbf{ComiCon}} & \multicolumn{3}{c||}{\textbf{Concert}} & \multicolumn{3}{c|}{\textbf{Helsinki}}  \\ \cline{2-10} 
\multicolumn{1}{c|}{\textbf{}}          & $\mathbf{\mu}$                     & \multicolumn{1}{c|}{$\tilde{\mathbf{x}}$} & \multicolumn{1}{c|}{$\mathbf{\sigma^2}$} & $\mathbf{\mu}$                     & \multicolumn{1}{c|}{$\tilde{\mathbf{x}}$} & \multicolumn{1}{c||}{$\mathbf{\sigma^2}$} & $\mathbf{\mu}$                     & \multicolumn{1}{c|}{$\tilde{\mathbf{x}}$} & \multicolumn{1}{c|}{$\mathbf{\sigma^2}$} \\ \hline
\multicolumn{1}{|l|}{\textbf{Baseline}} & \multicolumn{1}{r|}{.67} & .67         & .003 	& \multicolumn{1}{r|}{.75} 	& .75 	& \multicolumn{1}{r||}{.002} 	& \multicolumn{1}{r|}{.71} & .71 & .009 	\\ \hline
\multicolumn{1}{|l|}{\textbf{WFD-GM}} & \multicolumn{1}{r|}{.73} & .73           & .0006 & \multicolumn{1}{r|}{.84} & .86 & \multicolumn{1}{r||}{.003} & \multicolumn{1}{r|}{.71} & .70 & .008 	\\ \hline
\end{tabular}
\end{table}

We start discussing the two opposite scenarios in terms of nodes' mobility and geographical distribution: Concert and Helsinki.In the first one, as detailed in Tab.~\ref{tab:conn_components}, Baseline generated a highly fragmented $CG$ with more than 90 connected components, and the largest one containing only 2\% of the nodes. As we can note from Fig.~\ref{fig:concert_messages}, the message dissemination is really scarce with Baseline protocol due to the lack of nodes' mobility, which represents the only possibility of network reconfiguration for this protocol. In fact, in this case, groups are static until the end of the GOs resources (that go over the simulation time) and they are characterized only by 1-hop connections. In fact, as shown in Fig.~\ref{fig:concert_ccdf}, only 1\% of nodes are connected for the entire duration of the simulation, while all the others result not to be connected at all. Instead, with WFD-GM all the messages are disseminated in the network in the first 30 minutes of simulation, mainly thanks to merge and travelling operations (Fig.~\ref{fig:concert_messages}). All the nodes have a not null connection probability, even though their paths are characterized by a limited duration and, finally, the protocol provides a fully connected network over time (i.e., just one connected component with 100\% of the nodes).

On the other hand, in Helsinki scenario, Baseline and WFD-GM have similar performances. Specifically, in terms of messages dissemination (Fig.~\ref{fig:helsinki_messages}), the curves of the two protocols are mostly overlapped, reflecting the impact of the high nodes' mobility on the network reconfiguration and nodes' connectivity. In fact, in the first two hours, the percentage of the messages exchanged by nodes rapidly grows because the mobility model assumes that each node encounters several others on the way to the offices. Then, the messages dissemination slows down and stops for about 8 hours at approximately 80\%. After the working hours, curves rise again because most of the nodes move towards some meeting point (e.g., shopping center, restaurants, pubs), and this mobility supports the creation of new paths and connections with new nodes, exchanging thus new messages. Then, the curves become stable around 95\%. This can be due to the fact that part of the nodes came back to their homes position where there should be limited new connections. Looking at Tab.~\ref{tab:conn_components}, we can note that WFD-GM generates a less partitioned network than Baseline, even though both protocols create a large connected components including 99\% of the nodes, and Fig.~\ref{fig:helsinki_ccdf} shows a high connectivity probability for most of the nodes in both solutions. However, WFD-GM performs better than Baseline considering $T_D = 60$s.

ComiCon represents an intermediate scenario between the other two, both in terms of mobility and geographical area, highlighting significant advantages of WFD-GM with respect to Baseline in all the measures. In terms of message dissemination (Fig.~\ref{fig:comicon_messages}) it behaves as in the static scenario, thanks to merge and travelling operations and, even if Baseline performs better than in Concert scenario, it is not able to compete with WFD-GM. Then, Fig.~\ref{fig:comicon_ccdf} shows a total connectivity of the network for all $T_D$ parameters, much higher than Baseline, and this is also confirmed by the number of connected components in Tab.~\ref{tab:conn_components}.Therefore, we can summarize that WFD-GM improves the network connectivity and the message dissemination with respect to Baseline in scenarios characterized by medium and low mobility, and it performs similarly to Baseline in scenarios characterized by high mobility, since this represents a natural condition for WFD network reconfiguration. However, as shown in Tab.~\ref{tab:battery_stats}, it does not introduce additional overhead in terms of power consumption. In fact, it saves about 6\% of the devices' battery in ComiCon scenario, and it consumes about the same resources of Baseline in Helsinki scenario.

\section{Conclusions and future work}
\label{sec:conclusions}

In this work we present WFD-GM, a novel middleware-layer protocol for the autonomous configuration and management of Wi-Fi Direct groups in opportunistic networks, relying on commercial mobile devices.
WFD-GM is able to identify the best group configuration based on a context function that takes into account heterogeneous features of the devices (e.g., battery level, memory and CPU usage) and the current peers in proximity.
In addition, it enables inter-group communication by implementing groups' merge operations and enabling client nodes to ``travel'' between groups in proximity. In this way, they can contribute to messages dissemination relying on the classical store-carry-and-forward paradigm.
We validated WFD-GM through simulations of three realistic scenarios: a concert, a big convention, and a working day in an European city. We compared WFD-GM performances with a Baseline solution implementing a simplistic rule for the GO selection and not supporting additional network reconfiguration procedures.
In addition, we performed a set of real experiments to estimate the context parameters involved in the simulations (i.e., power consumption on commercial devices for WFD operations and maximum number of  incoming connections acceptable by a device acting as GO).
We showed that WFD-GM improves the network connectivity reducing the number of partitions and supporting higher connectivity probabilities for each pair of nodes in all the scenarios.
In addition, it improves messages dissemination with respect to Baseline in scenarios characterized by medium and low mobility, with a limited resource consumption.
It performs similarly to Baseline in scenarios characterized by high mobility, maintaining always a limited resource consumption.We are currently implementing a prototype of WFD-GM on Android devices to extend the protocol evaluation in real environments and, concurrently, we are investigating learning methods to allow WFD-GM to self-adapt its parameters to environmental changes and heterogeneous devices' characteristics.

\bibliographystyle{IEEEtran}
\bibliography{IEEEabrv,paper}

\end{document}